\documentclass[aps,pra,superscriptaddress,twocolumn]{revtex4-1}
\usepackage{dcolumn}
\usepackage[utf8]{inputenc}
\usepackage{amsmath}
\usepackage{amsfonts}
\usepackage{amssymb}
\usepackage{graphicx}
\usepackage{hyperref}
\usepackage{xcolor}

\usepackage{soul}

\newcommand{\ket}[1]{\ensuremath{\left|{#1}\right\rangle}}

\newcommand{\mb}{\mathbf}

\begin{document}

\title{Robust interferometric sensing using two-photon interference}

\author{G. H. Aguilar}
\affiliation{Instituto de F\'{\i}sica, Universidade Federal do Rio de Janeiro, Caixa Postal 68528, Rio de Janeiro, RJ 21941-972, Brazil}
\author{R. S. Piera}
\affiliation{Instituto de F\'{\i}sica, Universidade Federal do Rio de Janeiro, Caixa Postal 68528, Rio de Janeiro, RJ 21941-972, Brazil}
\author{P. L. Saldanha}
\affiliation{Departamento de F\'{\i}sica, Universidade Federal de Minas Gerais, 30161-970 Belo Horizonte-MG, Brazil}
\author{R. L. de Matos Filho}
\affiliation{Instituto de F\'{\i}sica, Universidade Federal do Rio de Janeiro, Caixa Postal 68528, Rio de Janeiro, RJ 21941-972, Brazil}
\author{S. P. Walborn}
\affiliation{Instituto de F\'{\i}sica, Universidade Federal do Rio de Janeiro, Caixa Postal 68528, Rio de Janeiro, RJ 21941-972, Brazil}
\affiliation{Departamento de F\'{\i}sica, Universidad de Concepci\'on, 160-C Concepci\'on, Chile}
\affiliation{Millennium Institute for Research in Optics, Universidad de Concepci\'on, 160-C Concepci\'on, Chile}
\begin{abstract}
Precise measurement of the angular deviation of an object  is a common task in science and technology. Many methods use light for this purpose.  Some of these exploit interference effects to achieve technological advantages, such as amplification effects, or simplified measurement devices. However, all of these schemes require phase stability to be useful.  Here we show theoretically and experimentally that this drawback can be lifted by utilizing two-photon interference, which is known to be less sensitive to phase fluctuations.  Our results show that non-classical interference can  provide a path towards robust interferometric sensing, allowing for increased metrological precision in the presence of phase noise.  
\end{abstract}

\pacs{05.45.Yv, 03.75.Lm, 42.65.Tg}
\maketitle

\section{Introduction}
Quantum metrology aims to employ quantum resources such as entanglement and squeezing to improve the estimation of physical parameters \cite{qmetro1,qmetro2}.   For example, it is well known that the use of entangled $N$-photon states can lead to a quantum advantage in interferometric sensing, allowing for Heisenberg scaling in the number of photons used to probe a phase parameter.  In this case the precision of estimation becomes proportional to $1/N$, in contrast to the shot-noise behavior ($\propto 1/\sqrt{N}$) that is attainable with $N$ independent photons. Though the focus in quantum metrology has typically been on surpassing the shot-noise limit, it may be possible to exploit quantum properties for other desirable advantages, such as improved dynamical range or robustness to noise \cite{lloyd08,tan08,weedbrook16,aguilar19,micadei15}.  
\par
 Monitoring of the tilt angle of an object is a metrological scenario that is relevant in several fields of science and technology \cite{kara1981,gillies,virgo,atom}.  A typical optical approach to this task  consists in tracking the spatial or phase displacements of a light beam reflected by the object \cite{park, howell,beamdeflection,alves2014weak,unbalancedweak,lyons16, alves2017,walborn18,walborn20a}.   Procedures based on monitoring a light beam that has reflected from the object,  illustrated in Fig. \ref{fig:example} a), involve the detection of the transverse spatial displacement of the beam, and thus require a detector that is position-sensitive, such as a camera or a quadrant detector.  Depending on the type of detector, the performance of such schemes can be degraded by distortions of the intensity distribution of the reflected light.  Phase displacement methods, which make use of interferometers such as the one displayed in Fig.~\ref{fig:example} b), can present advantages over  spatial-displacement procedures. In addition to providing novel interference features that can be exploited for the estimation of the tilt angle, such as weak-value amplification (WVA)~\cite{howell,beamdeflection,unbalancedweak}, these methods may reach levels of precision that  are not attainable via spatial-displacement-tracking schemes~\cite{park,walborn18,walborn20a}.   Besides their larger sensitivity,  some of these interferometric schemes require only  polarization measurements on the output light~\cite{alves2014weak,alves2017,walborn18,walborn20a}, allowing for the use of ``single pixel" detectors~\footnote{We use ``single pixel here to denote any detector without position sensitivity}.  
 \par
These interferometric techniques can suffer from a principal drawback.  Namely, nearly all interferometers are sensitive to unavoidable phase fluctuations.  Thus, it  is possible that  changes in the output light intensity that are due to variations of the physical parameter of interest (here $\theta$) cannot be distinguished from changes due to phase fluctuations caused by external effects, such as temperature or air currents, for example.  This not only decreases the precision of the parameter estimation, but can creates an ambiguity in the output signal. Moreover, if the characteristic timescale of the fluctuations is similar to that of the parameter of interest, it is not possible to use active phase control to stabilize these fluctuations, as this would suppress the sensitivity to changes in the parameter, as the two effects are indistinguishable. 
 \par
Here we show that the inherent disadvantage  of these interferometric methods can be overcome using two-photon interference.  When one photon of a pair probes an object, information about the motion of the object is imprinted on the two-photon state.  We show theoretically and experimentally that this information can be retrieved through simple coincidence counting in a two-photon Hong-Ou-Mandel interferometer \cite{hom87},  and furthermore that this technique can saturate the precision limits obtainable with single photons.  Moreover, the two-photon interference is not sensitive to fluctuations that change the optical path length by even a few wavelengths.   
Though the scheme we propose here does not increase the metrological precision when compared with single-photon interferometry in the ideal case, it does increase the robustness of the metrological protocol to phase instabilities, which is an effective increase in precision in the presence of phase noise.
\par
Our approach to stable interferometric sensing is reminiscent of the Hanbury Brown and Twiss (HBT) intensity interferometer developed in the 1950's \cite{hbt54}. This  novel interferometric technique used the correlation between the intensity registered by two separated detectors to measure the diameter of a light source, such as a distant star \cite{hbt56}.  This was an improvement on the Michelson stellar interferometer, which used standard interference of light (field correlations) for the same task.  Though it was successful in obtaining interference fringes between two beams even in poor atmospheric conditions \cite{michelson21},  the fringe pattern of the Michelson technique was not stable, but rather ``in motion", as it was sensitive to the relative phase between the two sources. The HBT interferometer, on the other hand, was not dependent on the relative phase, thus allowing for measurements with better resolution.  
\par
Though they seem like similar tasks, there is a distinct difference  between measuring the transverse diameter of a light source, and sensing changes in the tilt angle of a mirror, as considered in this manuscript.  In the Michelson stellar interferometer, the phase fluctuations can obscure the interference pattern, the period of which reveal information about the distance between the two point sources.  In the case we consider here, the phase changes can be actually indistinguishable from changes in the tilt angle ~$\theta$.
\par
From a quantum optics perspective, the Hanbury Brown and Twiss interferometer exploited the interference of two photons, each from a different source \cite{fano61}.  With the development of parametric down-conversion sources of photon pairs, and the seminal work of  Hong, Ou and Mandel (HOM) \cite{hom87,ouman89,Shih88}, two photon interference has found a wide range of applications, including synchronization of distant clocks \cite{Bahder04, quan16}, measurement of phase or amplitude distribution of photon pairs \cite{walborn03a,walborn04a,walborn05a,chen15}, Bell-state measurements for quantum information protocols \cite{mattle96,bouwmeester97,pan98,walborn03b,aguilar12,aguilar19} and optimal quantum cloning \cite{irvine04,nagali09}.  It is also at the heart of photonic quantum logic operations \cite{klm01,ralph02,pittman03}.   The present work is the application of two-photon interference towards robust optical sensing.     
 \par
  This article is organized as follows. In  section \ref{sec:problem}, we discuss the estimation of  tilt angle in more detail, showing how phase fluctuations can be quite detrimental in interferometric sensing of such deflections.  In section \ref{sec:HOMinterferometer}, we analyze the use of HOM interferometers for this metrological task and show that such devices are robust against phase instabilities.  Experimental results are presented in section \ref{sec:results}, where the stability of HOM measurements have been compared with a Sagnac interferometer. Conclusions  are provided in section \ref{sec:conclusions}.

\section{Measuring Angular Deflection}
\label{sec:problem}
\begin{figure}
\includegraphics[width=7cm]{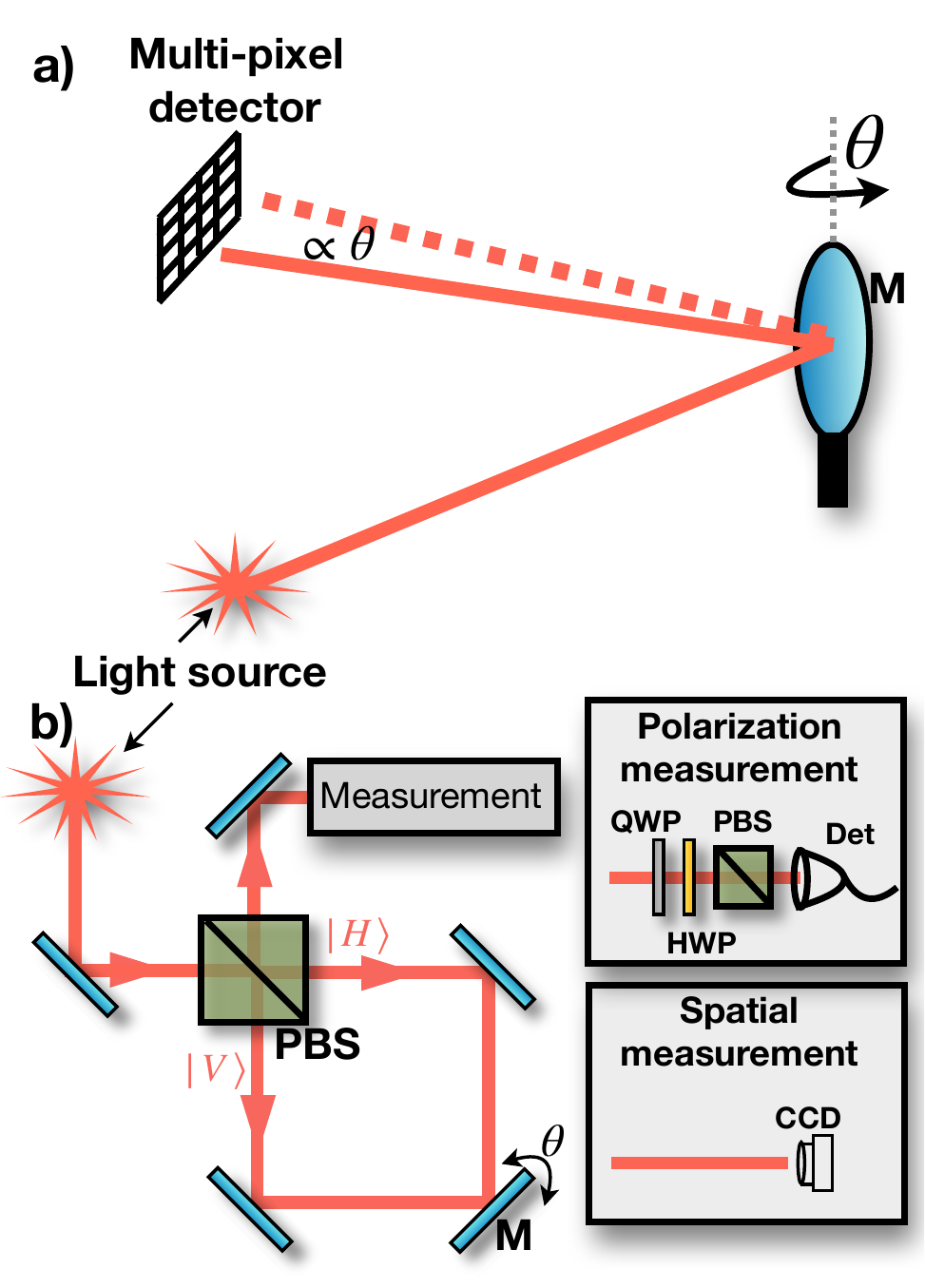}
\caption{a)  Basic spatial displacement scheme, in which the transverse spatial displacement of an optical field is measured to determine the angle $\theta$. b) Phase displacement scheme, in which the object is placed inside an interferometer, here constructed with a polarizing beam splitter (PBS).  The counter-propagating horizontal and vertical polarization components are displaced in opposite transverse directions.  Information about $\theta$ can be obtained by either measurements of the spatial degrees of freedom, accomplished by a CCD camera or quadrant detector, polarization measurements, implemented with wave plates, a polarizing beam splitter (PBS) and a single pixel detector, for example, or a combination of both.} 
\label{fig:example}
\end{figure}

Let us consider an optical field to monitor the angular mechanical motion of a (reflective) object, parametrized in this  example by the angle $\theta$.  A simple method is illustrated in Fig. \ref{fig:example} a), in which a light beam is sent towards the object and the transverse displacement of the reflected beam is measured.  If the transverse spatial profile of the beam is unaltered by the reflection from the object and the subsequent propagation, measurement of the  transverse displacement  is easily accomplished  with the use a quadrant detector, for example.  Otherwise, a camera must be used to obtain a reasonable amount of the transverse spatial distribution of the beam, in order to evaluate the displacement.  
  In this case, the ultimate precision limit for the estimation of $\theta$ with a beam of $\nu$  independent photons,  given by the Cram\'er-Rao bound \cite{lyons16,alves2017,walborn20a}, is
\begin{equation}
\Delta \theta \geq \frac{1}{\sqrt{\nu}4   k  \sigma}, 
\label{eq:QFI}
\end{equation}
where $\sigma$ is the width of the transverse profile of the beam  at the position of  the object and $k$ is the photon wavenumber.  Note that the here we have assumed that the photons sent to the object have a transverse spatial Gaussian distribution and that the transformation describing the tilting  of the object is unitary.
\par
As an alternative approach, a number of interferometric methods have been proposed and tested in the literature \cite{park, howell,beamdeflection,alves2014weak,unbalancedweak, alves2017,walborn18,walborn20a}.  In these schemes, the reflective object $M$ is placed inside an optical interferometer, playing the role of one of the mirrors.  Fig. \ref{fig:example} b) shows an example of a Sagnac interferometer constructed with a polarizing beam splitter (PBS), so that the horizontal ($H$) and vertical ($V$) polarization components follow counter-propagating directions. It is assumed that all optical devices are much more mechanically stable than $M$, so that any observed misalignment of the $H$ and $V$ beams can be attributed to the tilt angle $\theta$ of $M$. We note that the PBS can be replaced by an ordinary 50/50 beam splitter.  However, associating the counter-propagating paths to orthogonal polarization states allows for a simplified measurement scheme using waveplates and polarizers. It is well known that the Sagnac arrangement is one of the most phase-stable interferometers, since the $H$ and $V$ components follow the same optical path in different directions.   Such setups allow one to  take full advantage of the additional degree of freedom provided by the two paths of the interferometer.  As shown in Ref. \cite{alves2014weak,alves2017,walborn20a}, if the angular deviation of the object is small, nearly all of the relevant information about $\theta$ can be retrieved merely  through  polarization measurements on the output light.  In fact, the Sagnac, as well as other interferometric schemes, can saturate the precision limit \eqref{eq:QFI}. There is also the technical advantage that an area-integrating bucket (single pixel) detector can be used.    Additionally, the precision of the estimation of $\theta$ can be greatly boosted if the input beam enters the interferometer  slightly displaced from  the center of the input port~\cite{walborn18,walborn20a}.
\par
Interferometric techniques based on weak value amplification (WVA) have also been demonstrated \cite{howell,beamdeflection,unbalancedweak}.  In this case, in the context of Fig. \ref{fig:example} b), a polarizer selects a polarization state so that there is near-total destructive interference between the two beams, resulting in an output spatial distribution whose transverse displacement is amplified by a factor $A_w$.  The amplified displacement is measured using a quadrant detector or camera, as described above, and from this one can infer the deviation angle $\theta$.  While it  does not lead  to better precision in the estimation of $\theta$ \cite{jordan14,alves2014weak,zhang15}, when compared  to simple spatial displacement schemes, the WVA technique still  can be useful when the source intensity is much larger than the saturation threshold of the detectors, and can also provide improvements in the presence of some types of noise and detector saturation \cite{jordan14,viza15,harris17,Xu20}.   
\par

Though the interferometric  approaches can have several metrological advantages when compared to other techniques, they require phase stability in order to function properly. For example, if one considers obtaining information on the angle $\theta$ from the polarization degree of freedom alone, as in Refs. \cite{alves2014weak,alves2017,walborn20a}, the probability of obtaining one of two orthogonal polarization states  ($\pm$) when measuring the output light is of the general form:     
 \begin{equation}
 P_{\pm}(\theta,\varphi) = \frac{1}{2}\left(1 \pm \mathcal{V} f(\theta) \cos \varphi \right ),
 \label{eq:iprob_sagnac}
 \end{equation}
 where $\mathcal{V}$ is the visibility of the interference fringes, $f(\theta)$ is a real function related to the spatial profile of the beam and to the actual value of the angle $\theta$, whereas  $\varphi$ is the relative phase between the two paths of the interferometer. In principle, to obtain high-quality measurements of the angle $\theta$, it is sufficient that $\varphi$ remain approximately constant over the measurement acquisition time.   Thus, any observed variation of $ P_{\pm}(\theta, \varphi) $ can be attributed to changes in the angle $\theta$.   This is no longer the case if the phase $\varphi$ fluctuates.  Let us consider two scenarios.  In the first, we consider that the phase fluctuates around the reference phase on a time scale that is much shorter than the measurement time.  Let us assume that over the acquisition time, the phase can take values according to some probability distribution $Q(\varphi)$. In this case, the detection probability $P(\theta)$ is obtained by integrating $P_{\pm}(\theta,\varphi)Q(\varphi)$ over $\varphi$, obtaining $ P_{\pm}(\theta)=\frac{1}{2}\left(1\pm {\mathcal{V}}_Q f(\theta) \right )$,  where ${\mathcal{V}}_Q =\mathcal{V}\langle \cos \varphi \rangle_Q$ is the resulting visibility. Since $|\mathcal{V}_Q| \leq |\mathcal{V}|$, the effect of the phase fluctuations is to reduce the visibility (contrast of the interference fringes) and thus the measurement precision \cite{alves2017,walborn18}.  In many situations these effects can be partially mitigated through post-processing the observed data by filtering operations.  In the second scenario,  if $\varphi$ fluctuates on a timescale that is similar to that of the expected changes in the parameter $\theta$, then  one cannot distinguish between variations in $\theta$ or $\varphi$ by just monitoring $P_{\pm}(\theta,\varphi)$.  Moreover, one cannot filter the unwanted $\varphi$ information from the the data based on the characteristic time scale of the changes in $\theta$.  
 \par
\begin{figure}
\includegraphics[width=6cm]{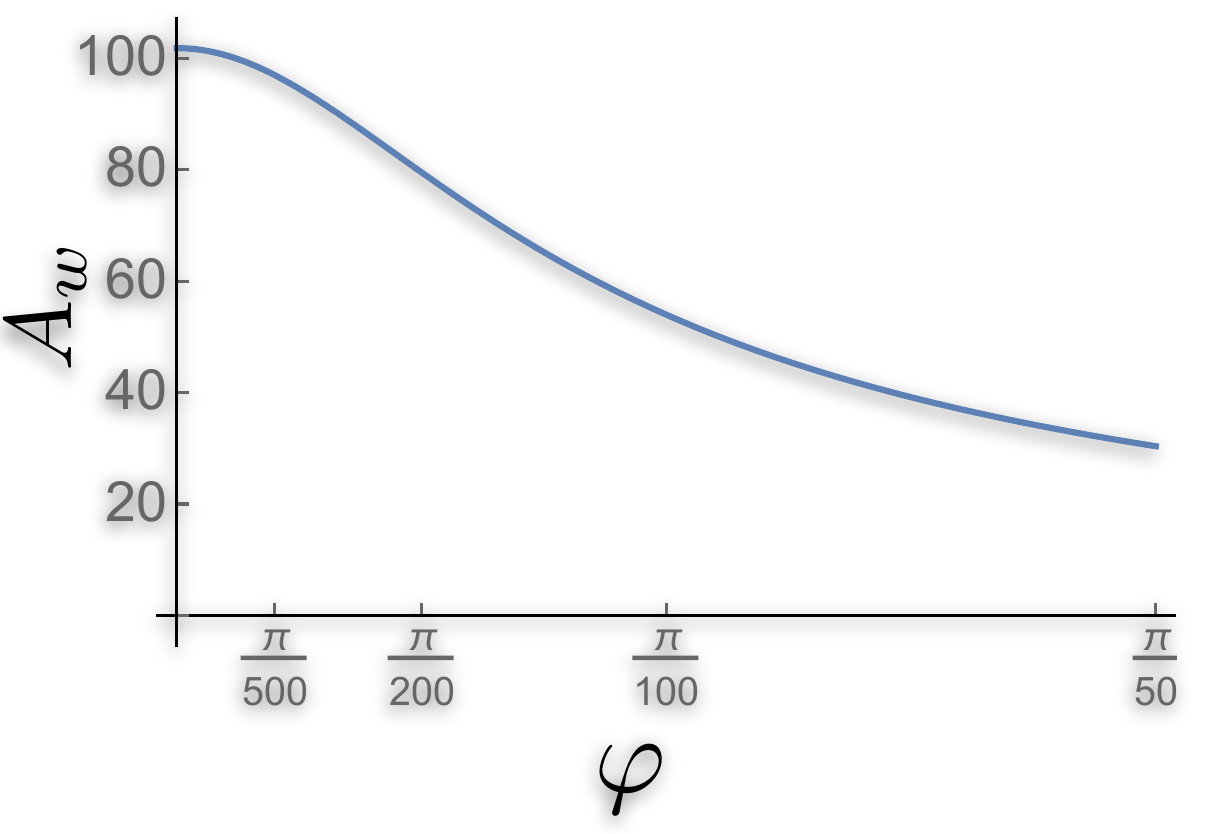}
\caption{Amplification factor in WVA measurement as a function of a phase shift $\varphi$.} 
\label{fig:weak-example}
\end{figure}
A similar effect appears in WVA techniques. For the interferometer shown in Fig. \ref{fig:example} b),  the amplification factor is given by  \cite{alves2014weak,alves2017}
 \begin{equation}
 A_w = \frac{{\langle \psi_{ps}|\hat\sigma_z| \psi_i\rangle}}{\langle \psi_{ps}| \psi_i\rangle},
 \end{equation}
 where $\ket{\psi_i}$ and $\ket{\psi_{ps}}$ are the initial and post-selected polarization states.  A phase fluctuation in the interferometer is essentially equivalent to changing the initial state \cite{Kedem12}, which will affect $A_w$, leading thus to erroneous results for the measurement of the beam displacement.  An example is shown in Fig. \ref{fig:weak-example} for $\ket{\psi_i} = (\ket{H} - e^{i \varphi} \ket{V})/\sqrt{2}$ and $\ket{\psi_{ps}}=\cos \theta\ket{H}+\sin\theta\ket{V}$, with $\theta=(1+1/80)\pi/4$.  One can see that a phase fluctuation of the order of $\pi/100$ reduces the amplification factor by roughly one-half,  severely affecting the estimation of $\theta$.      

 \section{Hong-Ou-Mandel Interferometer}
 \label{sec:HOMinterferometer}
  In contrast to single-photon interference, two-photon HOM interference~\cite{hom87} is insensitive to optical path length differences that are much smaller than the coherence length of the individual photons \cite{scully97}.  We show that one can exploit this property  in order to perform stable interferometric sensing of the tilt angle of an object. 
In the experimental setup depicted in Fig. \ref{fig:setup}a),  a HOM interferometer is built  with  a source of polarization-entangled photons and a polarizing beam splitter (PBS) \cite{aguilar12}.   The source is composed of two crossed-axes borate-barium (BBO) crystals  pumped by a $\lambda_p=$405 nm laser, producing degenerate pairs of photons at $\lambda=$810 nm, which we call signal ($s$) and idler ($i$), via spontaneous parametric down conversion (SPDC) \cite{kwiat99}. In the paraxial regime and in the thin crystals limit, the quantum state of the  photon pair at the near field of the crystals can be written as \cite{walborn03a} 
\begin{eqnarray}\nonumber
\ket{\Psi_0}=&&\int d\mb{q}_i\int d\mb{q}_s\Psi_p(\mb{q}_i,\mb{q}_s)\times \\
  &&\times\frac{1}{\sqrt{2}}\Big[\ket{\mb{q}_i,H}_i\ket{\mb{q}_s,H}_s+\ket{\mb{q}_i,V}_i\ket{\mb{q}_s,V}_s\Big],
\label{eq:init_state}
\end{eqnarray} 
where  $\mb{q}_j$ stands for  the transverse momentum  of each photon,  $\ket{\mb{q}_j,\epsilon}_j$  is a common eigenstate of transverse momentum and polarization of a single  photon ($V$ represents vertical and $H$ horizontal polarizations), and  $\Psi_p(\mb{q}_i,\mb{q}_s)=v(\mb{q}_i+\mb{q}_s)\gamma(\mb{q}_i-\mb{q}_s)$, where $v(\mb{q})$ is the angular spectrum of the  pump laser beam and $\gamma(\mb{q})$ is the phase matching function \cite{monken98a, walborn10,Schneeloch15}. The angular spectrum of the pump beam is transferred to the photon pair \cite{walborn10}, generating momentum entanglement, while the crossed-crystal geometry produces polarization entanglement \cite{kwiat99}.

Each photon of the pair is sent to a different input port of the PBS, which transmits (reflects) horizontal (vertical) polarized  photons.  Using single-mode fibers, the photons  at each output of the PBS are sent to avalanche-single-photon counters after polarization state projection (PP) implemented with a half-wave-plate (HWP) and another PBS before each detector. Coincidence counts are registered between pairs of detectors. The path length difference  between  the two arms of the interferometer is adjusted by moving the retroreflector in Fig. \ref{fig:setup}a).  We adjust  the path length such that  both photons arrive at the PBS with a time difference that is less than their coherence time, observing  HOM interference \cite{hom87}.  By setting PP to project the polarization state of the photons onto $\ket{++}$ ($\ket{+-}$), where $\ket{\pm}=1/\sqrt{2} (\ket{H}\pm\ket{V})$, destructive (constructive) interference is observed with a maximum HOM visibility of  96$\%$ (see Fig. \ref{fig:setup}b) and below). An advantage of using a PBS instead of a non-polarizing beam splitter is that there is always one photon in each exit port.  Thus, assuming that the classical $HH$ and $VV$ polarization correlations are preserved in propagation through the interferometer, all two-photon events can be registered in coincidence without the need for photon-number resolving detectors.
\subsection{Phase Stability}
\begin{figure}
\includegraphics[width=8.7cm]{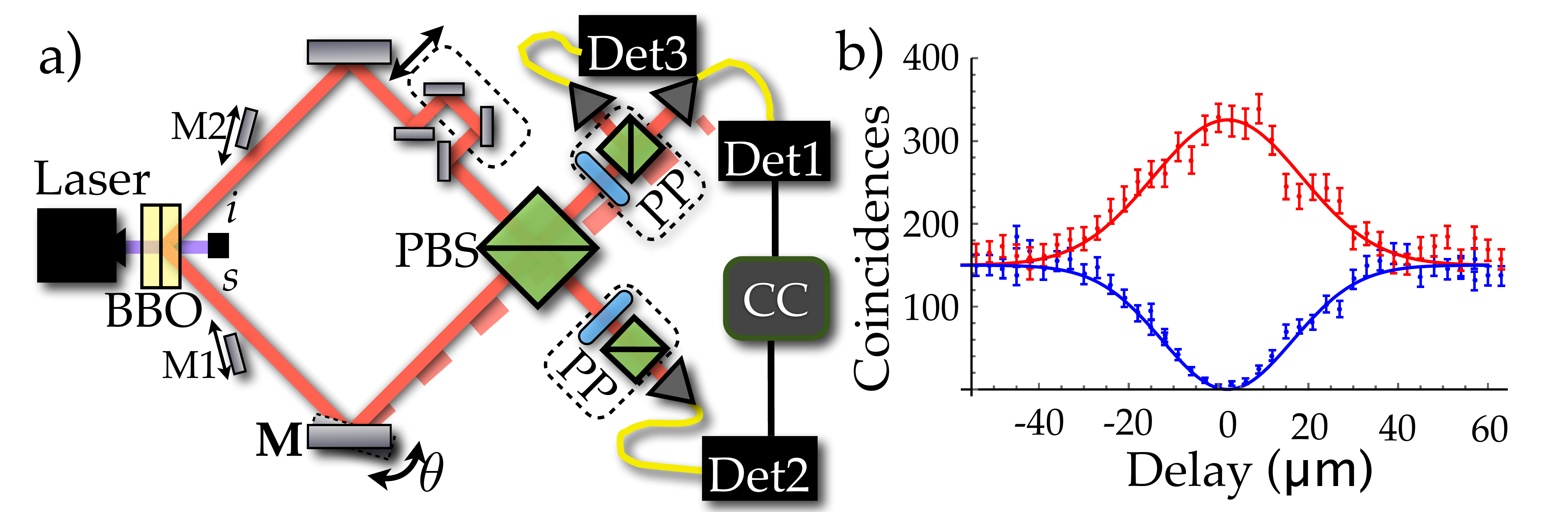}
\caption{a) Experimental Setup: An SPDC source creates pairs of entangled photons in the modes $\textit{i}$ and $\textit{s}$.  Photons in the mode $\textit{i}$ are reflected by a movable retroreflector and sent to a PBS. Photons in the mode $\textit{s}$ are sent to another input port of  the same PBS after being reflected by the mirror ${M}$, which has been tilted by an angle $\theta$. Photon pairs are detected in modes 1 and 2, where a polarization projection (PP) is implemented with a half-wave plate and a PBS. The photons are coupled into single-mode fibers, which are connected to single photons detectors whose output signal is analyzed with the CC electronics. Coincidences and single counts of the detectors are registered. Mirrors M1 and M2 are utilized to build a Sagnac interferometer. More details are provided in the main text. b) HOM interference observed when the path length traveled by photon $i$ is varied. Constructive and destructive interference is observe depending on the polarization projection at PP. Blue points correspond to coincidence counts, collected in 10 seconds, after polarization projection on the two-photon polarization state$\ket{+-}$, whereas red points correspond to coincidence counts after projection on state $\ket{++}$.}
\label{fig:setup}
\end{figure}

The HOM interference shown in Fig. \ref{fig:setup}b) can be characterized by the fitting function 
\begin{equation}
C(\delta) = A \left (1 \pm \mathcal{V}_z e^{-(2\pi\delta)^2/2 \Delta L^2} \right ), 
\end{equation} 
where $C(\delta)$ are the coincidence counts as a function of the path length difference $\delta$,  $\mathcal{V}_z$ is the visibility, and $\Delta L$ is the coherence length of the single photon fields.  The coherence length  $\Delta L \propto \lambda^2/\Delta \lambda$, where  $\lambda$ is the central wavelength of the down-converted light and $\Delta \lambda$ is the bandwidth obtained with the use of interference filters. For $\lambda = 810$nm and $\Delta \lambda = 10$nm, we have $\Delta L \sim 130 \lambda$, which agrees with the values obtained from the curve fit in Fig. \ref{fig:setup}b). 
Thus, in contrast to usual interferometry, changes in longitudinal path length on the order of a wavelength do not significantly affect the count rate in the HOM interferometer. This will be seen in detail in the following subsection.

\subsection{Tilt Angle Estimation}
 Considering the setup of Fig. \ref{fig:setup}a), let us now show that HOM interference can indeed be used to perform stable interferometric sensing of  the tilt angle $\theta$ of the reflective object $M$  with the same precision as is obtainable with  single-photon interferometry.  By design, we focus on a two-photon analog of the scheme discussed in section \ref{sec:problem} and illustrated in Fig. \ref{fig:example} b), since it saturates the precision limit \eqref{eq:QFI} for small tilt angles and does not require a detector that is sensitive to the intensity distribution (only polarization measurements are performed) \cite{alves2017,walborn18,walborn20a}.
In the following, the propagation direction of each photon will always be designated by $z$. This means that, in our notation,  we change the coordinate axes upon reflections, such that each photon propagation direction (when $M$  is not tilted) is defined along the positive $z$ axis even after the reflections. 
To simplify the notation, from now on we will represent only the $y$ component of the wave vector of each photon, denoting them by $q_i$ and $q_s$ for photons in mode $i$  and mode $s$, respectively, since the $x$ component is not affected by the tilting of mirror $M$. If the tilt angle $\theta$ is small, it changes the wave vector of the signal photon as $q_s\rightarrow q_s+2k\theta$, where $k=2\pi/\lambda$ is the photon wave number.  After propagation inside the interferometer, the quantum state of the photon pair at the  $z$-planes corresponding to detectors $1$ and $2$ (cf. Fig. \ref{fig:setup})  evolves from the state given in Eq. (\ref{eq:init_state}) to
\begin{eqnarray}
\ket{\Psi_1}&=&\frac{1}{\sqrt{2}}\int d{q}_i\int d{q}_s \Psi_p({q}_i,{q}_s) \Big[ e^{iq_{i}^2 z_{i,2}^{\rm tot}/2k}e^{iq_{s}^2 z_{s,M}/2k} \times\nonumber\\
              &\times& e^{i({q}_s+2k\theta)^2 z_{s,1}/2k} \ket{{q}_s+2k\theta,H}_1\ket{{q}_i,H}_2 \label{eq:state1}\\ \nonumber 
              &+&  e^{iq_{i}^2 z_{i,1}^{\rm tot}/2k} e^{iq_{s}^2 z_{s,M}/2k} e^{i({q}_s+2k\theta)^2 z_{s,2}/2k} \times \\ 
              \nonumber &\times&\ket{-{q}_i,V}_1\ket{-\left({q}_s+2k\theta\right),V}_2\Big],
\end{eqnarray} 
where modes 1 and 2 represent the two output modes of the central PBS of Fig. \ref{fig:setup} a), and $z_{i,1}^{\rm tot}$ ($z_{i,2}^{\rm tot}$) is the optical path length (OPL) of the photon in mode $i$, when propagating from the crystal to detector $1$ ($2$).   $z_{s,M}$ is the OPL of the photon in mode $s$, when propagating from crystal to mirror $M$, and $z_{s,1}$ ($z_{s,2}$) is the OPL of photon $s$ when propagating from mirror $M$ to detector $1$ ($2$).  Note that a reflection inverts the $y$ component of the wave vector.    

Before reaching detectors $1$ and $2$,  the polarization state of the photon pair is projected (using PP) onto the polarization basis $\{\ket{++},\ket{--},\ket{+-},\ket{-+}\}$, where  $\ket{\pm}$ are linear diagonal polarization states of a single photon.  If detectors $1$ and $2$ are point detectors, the probability of a coincidence count at coordinates ${q}_1$ and $q_2$ in momentum space is  $P^{(\pm)}(q_1,q_2,\theta) = |\langle q_1, \pm| \langle q_2,\pm|\Psi_1\rangle|^2$, where the ``$(\pm)$" notation in the superscript refers to cases when the photons are detected with the same ($++$ or $--$) or different ($+-$ or $-+$) linear polarization states.  Using state \eqref{eq:state1}, we have in this case
\begin{widetext}
\begin{align}
	P^\pm(q_1,q_2,\theta)&=  \frac{1}{4}\left\{|\Psi_p(q_2, q_1-2k\theta)|^2 +|\Psi_p(-q_1,-q_2-2k\theta)|^2\right. \nonumber \\ 
	& \pm2|\Psi_p(q_2,q_1-2k\theta)||\Psi_p(-q_1,-q_2-2k\theta)|  
	\times\left.\cos [ (q_1^2-q_2^2)\frac{\Delta z}{2k}+ 2 (q_1+q_2)z_{s,M}\theta +\Phi]\right\},
\label{eq:P2}							
\end{align}\end{widetext}
where $\Phi=\arg[\Psi_p(-q_1,-q_2-2k\theta)]-\arg[\Psi_p(q_2,q_1-2k\theta)]$,  $\Delta z=z_i^{\rm tot}-z_s^{\rm tot}$, $z_s^{\rm tot}=z_{s,M}+z_s$, and we have set $z_{i,1}^{\rm tot}=z_{i,2}^{\rm tot}=z_i^{\rm tot}$ and $z_{s,1}=z_{s,2}=z_s$. 
The first term in the argument of the cosine function arises from the phase curvature of the down-converted beams, and does not depend on $\theta$. It is responsible for a momentum-dependent phase shift that is controlled by  the overall difference of OPL between  the two arms of the interferometer. In the paraxial regime and for degenerated photon pairs, one has for each photon $q^2\ll k^2$ and $k_z\approx k\left(1-q^2/2k^2\right)$.  This implies that $(q_1^2-q_2^2)/2k\approx k_{z_1}-k_{z_2}$ is much  smaller than $k$ for all relevant values of $q_1$ and $q_2$. This is the well-known phase stability of two-photon interference \cite{scully97}:  even if the difference $\Delta z$ of  OPL between the two arms of the interferometer changes by several wavelengths, the corresponding change in $P^\pm(q_1,q_2,\theta)$ will be negligible.  If detectors $1$ and $2$ are area-integrating bucket detectors (see below), coincidence counts will correspond to integrating  $P^\pm(q_1,q_2,\theta)$ over the momenta. The $\Delta z$ term will then be averaged and will degrade the visibility of $P^\pm(\theta)$, although this effect will be negligible for realistic values of $\Delta z$.
It is illustrative to compare this to a similar calculation using a single-photon interferometer, which results in an equation similar to \eqref{eq:P2}, with a modified term  $(k-q^2/2k)\Delta z\approx k_z\Delta z$ replacing the first term in the argument of the cosine function. Since the magnitude of $k_z$  is approximately equal to the magnitude of $k$, a variation of $\Delta z$ over a fraction of the photon wavelength will considerably change the value of  $P_{1}^\pm(q,\theta)$, in stark contrast to the HOM interferometer case. If an area-integrating bucket detector is used, the visibility of $P_1^\pm(\theta)$ will  be sensitive to fluctuations of $\Delta z$ that are as small as a fraction of the photon wavelength. In this regard the HOM interferometer is much more robust to fluctuations of the OPL diference between the interferometer arms than a single-photon interferometer.

The transverse spatial amplitude of the photons pairs can be approximated by a product of two Gaussians \cite{Schneeloch15}. In this case, the transverse profile of the pump beam and the phase matching function are given by Gaussian functions
\begin{eqnarray}
v({q})&=&A\mathrm{exp}\left[\frac{-q^2w_p^2}{4}\right], \\ \nonumber
\gamma({q})&=&B\mathrm{exp}\left[\frac{-q^2\omega_c^2}{4}\right],
\label{eq:doublegauss}
\end{eqnarray}
where $w_p$ is the transverse width of the gaussian pump beam,  $\omega_c \sim L/4k_p$ is the characteristic width of the phase matching function,  $L$ is the length of the non-linear crystals, and $k_p$ the wavenumber of the pump beam. $A$ and $B$ are real normalization coefficients.  Under this approximation, the coincidence count probability is  given  by

 \begin{eqnarray}
	&& P^\pm(q_1,q_2,\theta)=  \frac{1}{2}A^2B^2v^2(q_+)\gamma^2(q_--2k\theta) e^{-2w_p^2k^2\theta^2} \times  \nonumber\\
	& & \left \{ \cosh \left[2q_+w_p^2 k\theta\right] \pm \cos \left[\frac{q_+q_-}{2k}\Delta z+2 q_+z_{s,M}\theta \right] \right \},			\label{eq:P3}\end{eqnarray}
where we introduce  $q_{\pm}=q_1 \pm q_2$.

 Let us  assume now that detectors $1$ and $2$ are area-integrating bucket detectors that measure solely the joint probability of projecting the output beams on the same polarization state ($+$) or on ortogonal polarization states ($-$). In this case, the coincidence counts give information on the probabilities $ P^\pm(q_1,q_2,\theta)$ integrated over all values of $q_1$ and $q_2$, resulting in
\begin{widetext}\begin{equation}
P^\pm(\theta)= \frac{1}{2}\left[1\pm  \mathcal{V}_z \exp\left\{-2\left[w_p^2 k^2+\frac{1}{w_p^2}\left(1-\frac{\Delta z^2}{4k^2\omega_c^2w_p^2}\right)\left(\frac{\Delta z}{2}+z_{s,M}\right)\right]\theta^2\right\}\right],
\label{eq:P44}
\end{equation}\end{widetext}
where $ \mathcal{V}_z=1-\Delta z^2/(8k^2\omega_c^2w_p^2)$ is the visibility of $P^\pm(\theta)$ and we have assumed that $\Delta z/2k\ll\omega_cw_p$.
As discussed above,  even for values of $\Delta z$ as large as several wavelengths of the degenerate photons, the visibility $\mathcal{V}_z$ will be extremely close to unity.

When $\Delta z=0$, that is, when the difference of OPL between the two arms of the interferometer  vanishes,  $P_\pm(\theta)$ reduces to
\begin{eqnarray}
P^\pm(\theta)&=& \frac{1}{2}\left(1\pm e^{-2 \left[k^2w_p^2+z_{s,M}^2/w_p^2\right]\theta^2}\right). \label{eq:P4}\\ \nonumber
&=&\frac{1}{2}\left(1\pm e^{-\frac{1}{2}k_p^2w_p^2(z_{s,M})\theta^2}\right),
\end{eqnarray}
where $k_p=2k$ is the wavenumber of the pump laser and $w_p(z_{s,M})$ is the width of the pump laser at the mirror $M$. This expression coincides precisely with the probability of projecting the output light of a perfectly aligned single-photon interferometer on the diagonal polarization basis $\{\ket{+},\ket{-}\}$,  when the width of the beam at object $M$ is $w_p(z_{s,M})$ and the wavenumber of the photons is equal to $k=k_p/2$~\cite{walborn20a}. The fact that we obtain exactly the same result in both setups shows that a perfectly aligned HOM interferometer can reach the same precision in the estimation of a tilt angle $\theta$ as a perfectly aligned single-photon interferometer, which is known to saturate  the precision limit \eqref{eq:QFI}.  Of course, here there is an extra overhead of a factor of 2, since HOM interference requires a photon pair instead of a single photon.  Still, this allows us to unambiguously attribute the change in coincidence counts to variation of $\theta$, even in the presence of OPL fluctuations.

\section{Experimental Results}
\label{sec:results}

\begin{figure}
\includegraphics[width=8.5cm]{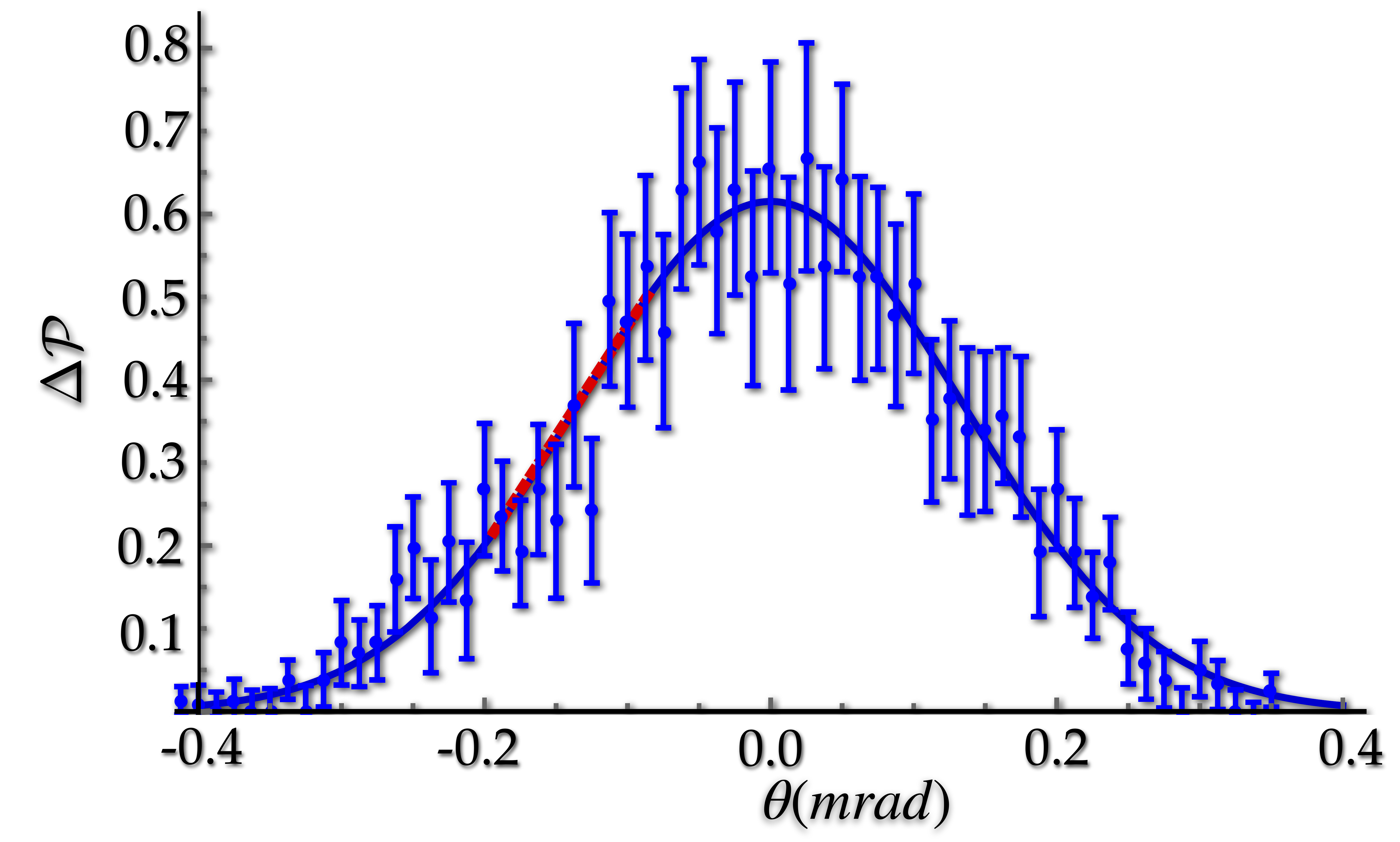}
\caption{Difference of the normalized counts $\mathcal{P}^\pm(\theta)$ as a function of the tilting angle $\theta$ . The  dots correspond  to  measured values, while the solid curve is a Gaussian fitting of the experimental data.  The red dashed line is a linear fitting around the point of maximal variation of the probabilities with $\theta$. }
\label{fig:tilting_probs}
\end{figure}

In order to experimentally confirm the analysis  presented in section~\ref{sec:HOMinterferometer}, we have implemented the experimental setup shown in  Fig.~\ref{fig:setup}a)  and described at the beginning of section~\ref{sec:HOMinterferometer}. We measured the behavior of normalized coincidence counts $\mathcal{P}^{\pm}(\theta)$ while the angle of mirror $M$ in \ref{fig:setup}a) is varied. We compare the results with the probability  $P^\pm(\theta)$ (see Eq.~\eqref{eq:P4}) of projecting the output photons onto the same polarization  state or onto orthogonal states. Coincidence counts  for each polarization setting were registered. Here $C_{++}$ ($C_{+-}$) refers the coincidence counts when the photons have been projected onto the same (orthogonal) polarization states using PP.   We determine  $\mathcal{P}^{\pm}(\theta)=C_{+\pm}(\theta)/(C_{++}^{max}+C_{+-}^{max})$, where $C_{++}^{max}$ are the coincidence counts when $\theta=0$.  The difference $\Delta{\mathcal{P}}  = \mathcal{P}^+(\theta)- \mathcal{P}^-(\theta)$ is  shown in Fig.~\ref{fig:tilting_probs} and reproduces the behavior expected from Eq.~\eqref{eq:P4}.  This confirms that polarization measurements in the output photons of a HOM interferometer do indeed give information about the tilt angle of object $M$. By fitting the experimental points with a Gaussian function, we find a standard deviation of $129(5) \mu$rad, which is in rough agreement with the theoretical value of $1/(2 k w_p)\approx133  \mu$rad from Eq. \eqref{eq:P3}, with $\lambda=810$nm,  $w_p \approx 0.5 \times 10^{-3}$mm.   The red dotted line shows the region of maximum sensitivity \cite{walborn18}, and has a slope of $\sim 2.65$mrad$^{-1}$.  Thus, in a sensing application, $M$ should be initially aligned at a value $\theta_0$ in this region, so that deviations $\theta$ with the best precision can be obtained using the relation $\theta \approx 0.38 \Delta{\mathcal{P}}$mrad.

 To study the stability of the estimation of tilt angle based on HOM interferometry, we performed measurements for a period of 8 hours  and compare the results with a similar measurement realized with a  Sagnac interferometer under the same conditions. The Sagnac interferometer is known to be the most stable single-photon interferometer, since the light beams propagate in opposite directions through  the same set of mirrors, resulting in an interferometer with high stability against mechanical fluctuations. Moreover, it has been shown recently that such an interferometer can saturate the ultimate precision limits in the estimation of small tilt angles~\cite{alves2017,walborn18,walborn20a}.  To make a fair comparison between the two interferometers, the Sagnac interferometer was constructed from nearly the same optical arrangement as the HOM interferometer.   By  simply  inserting the mirrors M1 and M2 to the setup of Fig.~\ref{fig:setup}a) and using the fiber-coupler of detector 2 to back-inject an attenuated laser beam, a common-path Sagnac interferometer  was built, where the $H$ and $V$ polarization components traversed nearly the same paths as the down-converted photons.

 \par
\begin{figure}
\includegraphics[width=8.5cm]{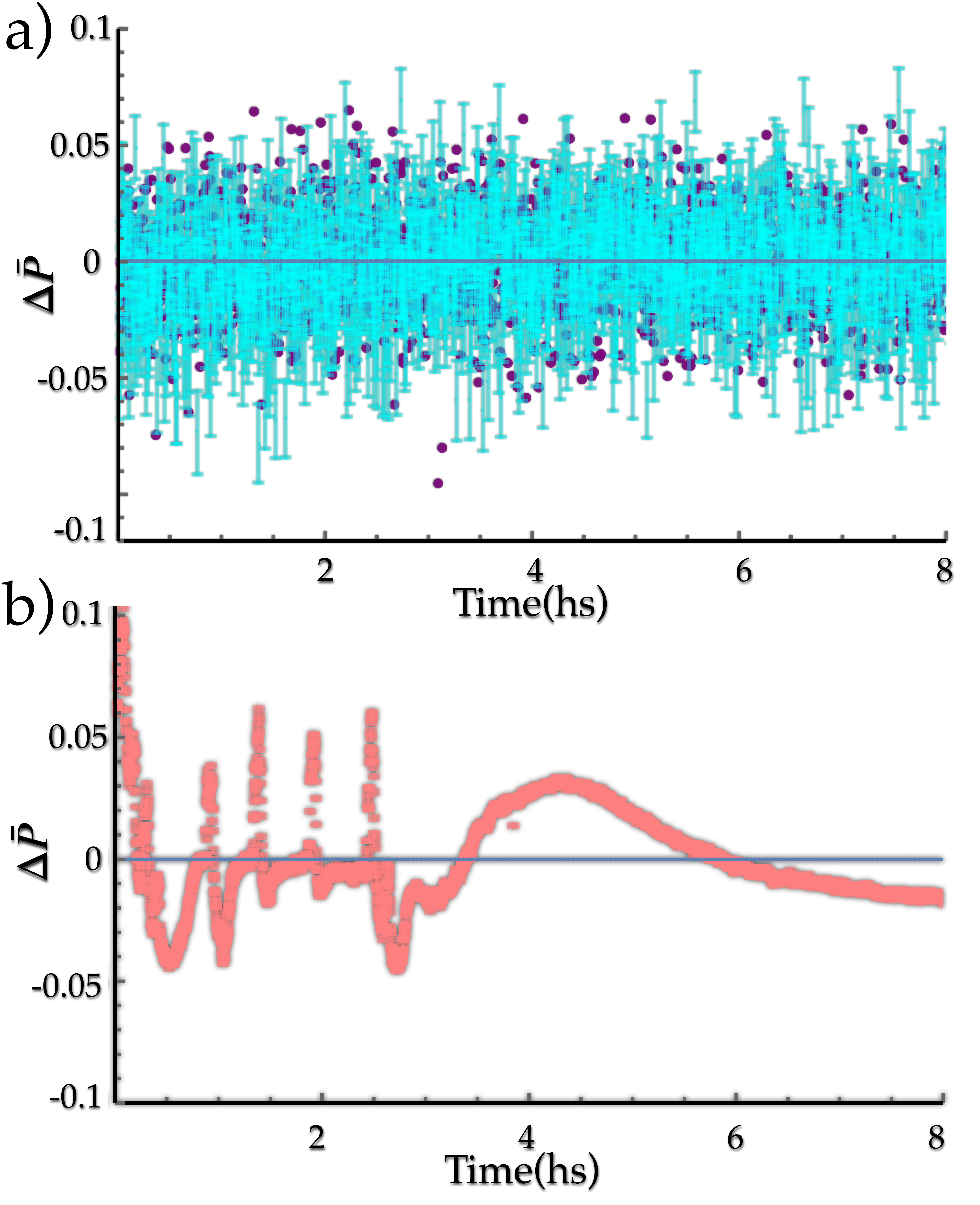}
\caption{The upper figure shows in cyan the temporal fluctuation $\Delta \bar{P}$ of the the probabilities difference $P^+(\theta)-P^-(\theta)$ for the HOM interferometer  around its mean value over a period of 8 hours. The purple dots correspond to numerically generated values obeying Poissonian statistics, showing that the observed fluctuations are due to the low count regime. The lower figure shows the corresponding quantity for the Sagnac interferometer.}
\label{fig:stability}
\end{figure}

 In both cases we tuned the parameters of each interferometer  to the most sensitive region, given by the maximum derivative of   Eq. \eqref{eq:P4} and represented by the red line in Fig. \ref{fig:tilting_probs}.   In Fig. \ref{fig:stability}a), we show the difference $\Delta \bar{P} = \Delta P - \mathrm{Mean}(\Delta{P})$ as a function of the time for the HOM interferometer, where $\mathrm{Mean}(\Delta{P})$ is evaluated over all 8 hours of data. The corresponding result for the Sagnac interferometer is shown in Fig.~\ref{fig:stability}b),  where the probabilities $P_{\rm Sag}^\pm(\theta)$ were determined after single counts in detectors 1 and 3 were registered ($\sim 80,000$ total counts/s). Notice that in the absence of optical path length fluctuations both signals should remain constant at 0, except for higher-frequency noise stemming from the statistics of photon counting and from the non-ideal quantum efficiency of the detectors. The large amount of noise  in the signal from the HOM interferometer, when compared to the signal from the Sagnac interferometer,  comes from the very low coincidence count rate ($\sim 25$ total coincident counts/s), which leads to a higher amount of relative noise ($\sim 1/\sqrt{\nu}$) in the photon count statistics, where $\nu$ here is the number of events. To verify that the noise in Fig. \ref{fig:stability}a) comes exclusively from the low coincidence rate, we  numerically simulated coincidence counts considering random Poissonian count statistics.  The results of this simulation are plotted as purple dots in \ref{fig:stability}a). One can observe that there is a full overlap of the simulated and  the experimental data, showing that the large noise comes from the very low coincidence count rate. To increase coincidence count rates, one can  increase the power of the laser pumping the SPDC crystal and improve the non-linear source, optical coupling, and efficiency of the detectors \cite{Kurtsiefer01,Bovino2020}.  To achieve the same count statistics in both setups one would need to increase the acquisition time for the HOM interferometer, due to the fact that coincidence count rates are always lower than single count rates since one must detect two photons.  This is thus a limiting factor for HOM interferometry, and is relevant in the case of fast phase variations. If the overall single-photon detection efficiency (including optical losses and detection efficiency) are $\eta$, then the overall two-photon efficiency is roughly $\eta^2$, where $0\leq \eta \leq1$.  Thus, the precision provided by the HOM interference must be better than that of the single-photon interference by a factor $\eta$ in order to be advantageous when losses and efficiencies are taken into account. 
 
\begin{figure}
\includegraphics[width=8.5cm]{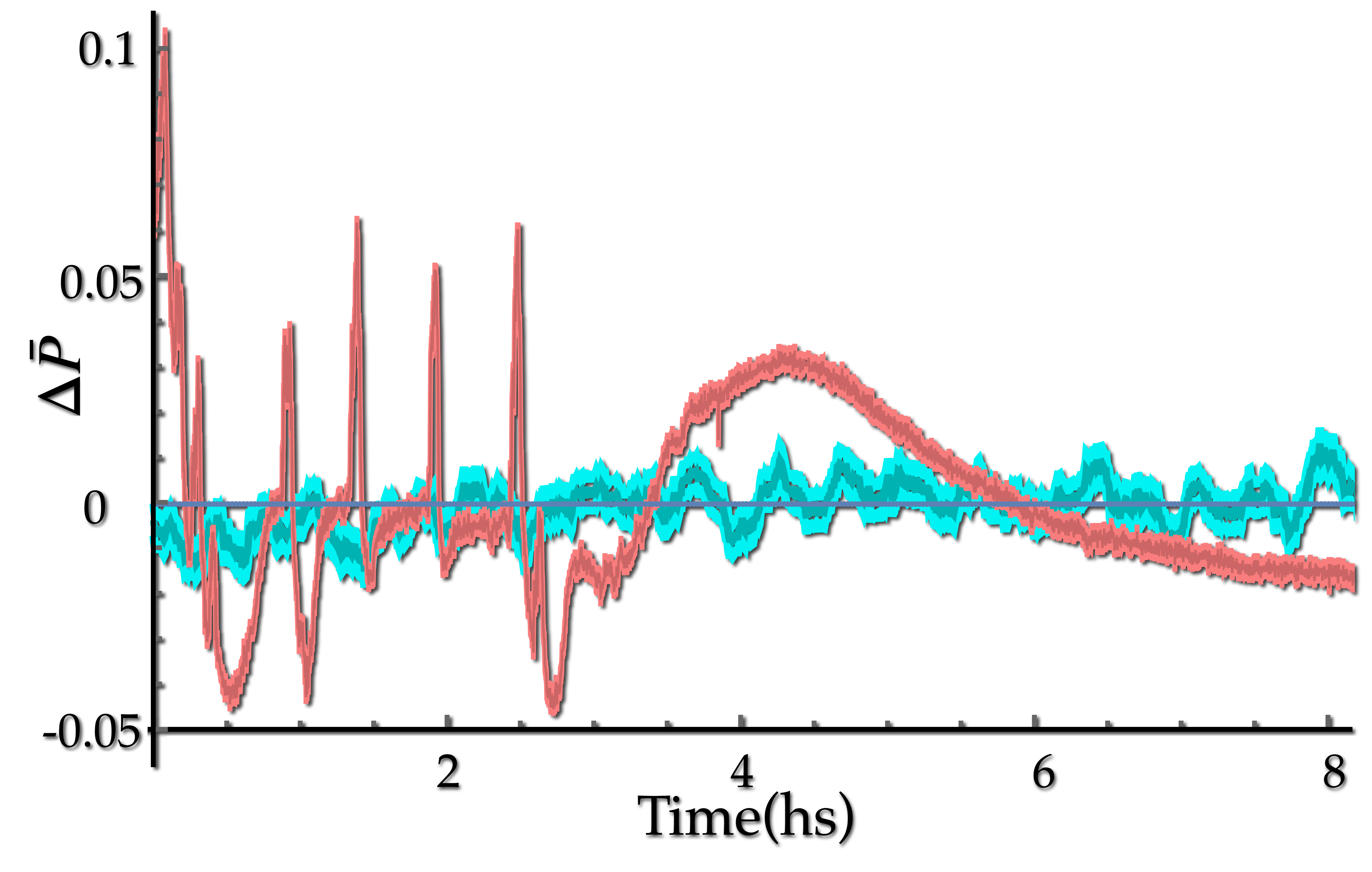}
\caption{Temporal fluctuation $\Delta \bar{P}$ of the the probabilities difference $P^+(\theta)-P^-(\theta)$ around their mean values over a period of 8 hours.  The cyan curve shows the behavior for the HOM interferometer after a mean filter has been applied to it to eliminate photon count uncertainty, while the red curve shows the behavior for the Sagnac interferometer.  The blue line corresponds to the situation of perfect stability of the interferometers. The shadowed regions around the curves correspond to the error bars}
\label{fig:stabilityfin}
\end{figure}

 For the case of the Sagnac interferometer, with relative noise $\sim 10^{-3}$, variations of $\Delta \bar{P}$ are observed due to phase fluctuations and drift in the interval from 0 to 4 hours.  After that, the counts have a much slower variation around the mean value.  To better compare the temporal stability of both signals, we processed the HOM coincidence counts in order to eliminate the noise from low photon counting statistics (relative noise $\sim 10^{-1}$).  For this, we take intervals of 10 minutes and calculate the mean values of $\Delta \bar{P}$ in these intervals (relative noise $\sim 10^{-2}$).   The result is shown in Fig.~\ref{fig:stabilityfin}. Here, we can observe that the phase fluctuations in Sagnac measurements generate variations in $\Delta \bar{P}$ that are roughly an order of magnitude larger than the variations in HOM interferometer.    This clearly demonstrates that in this metrological scenario the HOM interferometer is much more robust against OPL fluctuations than the Sagnac interferometer, without any active stabilization. Moreover, all interferometers must be initially aligned to an initial reference phase. At some later time, it is typically necessary to detect and correct phase drift, realigning the relative phase to the initial value \cite{Oh2020}. This frequency at which this process must be realized  is related to the inherent phase instability of the interferometer. Given that the HOM interferometer is stable over time scales that are much larger than that of single-photon interferometers, they have the advantage that the realignment procedure can be much less frequent.  In addition, we note that the stability of the two-photon interference depends upon the coherence length of the down-converted fields, which can be made large by decreasing the bandwidth of the down-converted light.
 \par
 We note that the effect of OPL fluctuations in the Sagnac interferometer is not a question of reduced precision, as would occur for variations  on a time scale that is much shorter than the detection time (8 s).  Rather, here one cannot determine whether the observed change in optical signal is due to variation of the tilt angle $\theta$, or simply an OPL fluctuation.   To the best of our knowledge, the standard active stabilization techniques using auxiliary lasers are unable to control the unknown OPL fluctuations, since the exact same problem 
 will exist for these beams.

\section{Conclusions}
\label{sec:conclusions}
We have  shown, both theoretically and experimentally, that two-photon HOM interferometry can be useful to counteract the effects of phase instability in metrological protocols based on optical interferometry. We demonstrate that the coincidence count probabilities associated to simple polarization measurements reproduce the detection probabilities in common metrological schemes using single-photon interference, which are known to saturate the ultimate precision limits.  Thus, in the absence of fluctuations, the use of two photon interference does not lead to an enhancement in the metrological precision when compared to ideal single-photon interferometric setups.  On the other hand, in non-ideal situations, where the optical path length can vary indeterminately due to temperature or air currents, the robustness of two-photon interference can lead to better precision in the interferometric estimation of physical parameters.  This indicates that quantum effects, besides being useful for enhancement in precision due to Heisenberg scaling,  can offer additional metrological advantages. We provide an experimental comparison between a HOM interferometer and a Sagnac interferometer over 8 hours of operation, clearly displaying the robustness  of a HOM interferometer against phase fluctuations even without the use of active stabilization. Moreover, our results suggest that with the use of two-photon interference, active phase stabilization is not necessary.  We expect our results to motivate the search for new multi-photon interferometric techniques to optimize noisy optical metrology.  One interesting feature to be explored in the future is to consider the effect of losses and how to mitigate them.  In this regard, we note that when the idler photon is lost, the signal photon still contains information on theta, since its spatial degrees of freedom have suffered the angular deviation from $\theta$. This information could be recovered in the single counts in principle via a spatial measurement.


\begin{acknowledgements}
The authors would like to thank CAPES, CNPQ and the INCT-IQ for partial
financial support.  This work was realized as part of the CAPES/PROCAD program.  SPW received support from the Chilean Fondo Nacional de Desarrollo Cient\'{i}fico y Tecnol\'{o}gico (FONDECYT) (1200266) and the Millennium Institute for Research in Optics (MIRO). 

\end{acknowledgements}

\bibliographystyle{apsrev}

\end{document}